# Observation of Flat Band and Van Hove Singularity in Non-superconducting Nitrogen-doped Lutetium Hydride

X. Liang[1,†], J. Zhang[2,†], Z.-H. Lin[1,†], J.-F. Zhao[2,†], S.-Y. Feng[1], W.-L. Lu[1], G.-D. Wang[2], L.-C. Shi[2], N.-N. Wang[2], P.-F. Shan[2], Z. Zhang[1], M. Naamneh[3], R.-Z. Liu[4], B. Michon[1,5], J.-G. Cheng[2], C.-Q. Jin[2], Y. Ren[1,*], J.-Z. Ma[1,5,6,*]

[1]*Department of Physics, City University of Hong Kong, Kowloon, Hong Kong, China*
[2]*Beijing National Laboratory for Condensed Matter Physics and Institute of Physics, Chinese Academy of Sciences, Beijing, 100190 China*
[3]*Department of Physics, Ben-Gurion University of the Negev, Beer-Sheva, 84105, Israel*
[4]*Institute of Molecular Sciences and Engineering, Institute of Frontier and Interdisciplinary Science, Shandong University, Qingdao 266237, P. R. China.*
[5]*Hong Kong Institute for Advanced Study, City University of Hong Kong, Kowloon, Hong Kong, China*
[6]*City University of Hong Kong Shenzhen Research Institute, Shenzhen, China*

† *The authors contributed equally to this work.*
\*Corresponding to: Junzhang Ma (junzhama@cityu.edu.hk), Yang Ren (yangren@cityu.edu.hk)

**Hydrogen-rich materials offer a compelling avenue towards room temperature superconductivity, albeit under ultra-high pressure. However, the experimental investigation of the electronic band structure remains elusive, due to the inherent instability of most of the hydrogen-rich materials upon pressure release. Very recently, nitrogen-doped lutetium hydride was claimed to host room temperature superconductivity under near ambient pressure but was disproven by following works. Upon decompression, nitrogen doped lutetium hydride manifests a stable metallic phase with dark blue color. Moreover, high temperature superconductivity has been reported in lutetium hydrides $Lu_4H_{23}$ (~71 K) under around 200 GPa. These properties engender an unprecedented opportunity, allowing for the experimental investigation of the electronic band structure intrinsic to hydrogen-rich material. In this work, using angle resolved photoemission spectroscopy to investigate the non-superconducting nitrogen doped lutetium hydride, we observed significant flat band**

**and Van Hove singularity marginally below the Fermi level. These salient features, identified as critical elements, proffer potential amplifiers for the realization of heightened superconductivity, as evidenced by prior research. Our results not only unveil a confluence of potent strong correlation effects and anisotropy within the Lu-H-N compound, but also provide a prospect for engineering high temperature superconductivity through the strategic manipulation of flat band and the VHS, effectively tailoring their alignment with the Fermi energy.**

The realization of room temperature superconductivity stands as the paramount objective in the realm of condensed matter physics, carrying profound implications for industrial revolutionary advancements. Hydrogen-rich materials emerge as a focal point in the pursuit of high temperature superconductivity[1,2], exemplified by the manifestation of superconductivity in many compounds under high pressure, such as in $H_3S$ and $LaH_{10}$ [3–7], yielding critical temperatures higher than 200K. Theoretical analysis predicts the prospective attainment of room temperature superconductivity in $YH_{10}$ achievable under extreme pressures of megabars[6]. The elucidation of the electronic band structure assumes a pivotal role in comprehending the mechanism of the superconductivity. However, the high-pressure environment precludes direct experimental investigation of the band structure of hydrogen-rich materials. Even under ambient pressure, the prospect of scrutinizing said band structure remains elusive, due to the inherent instability of most hydrogen-rich materials upon pressure release. Very recently, room temperature superconductivity was experimentally claimed in nitrogen doped lutetium hydride (Lu-H-N) at near-ambient pressure[8]. Notably, ensuing discourse has ensued, prompted by subsequent works asserting the absence of superconductivity within the same compound[9–12]. Later, high temperature superconductivity with critical temperature up to 71K is reported in lutetium polyhydrides[13]. Meanwhile, material Lu-H-N exhibits a stable metallic phase upon pressure relaxation[8] and calculation shows that the band structure changes little versus different pressure[14], thereby affording scientists a unique avenue to embark upon empirical investigation of the electronic band structure experimentally at ambient environment but reflecting the pressured situation. In the exploration of hydride-rich superconductors, the quest for strategies to elevate the critical temperature and concurrently mitigate pressure constitutes a pivotal line of inquiry.

On one facet, the intriguing concept of the flat band, delineated by their negligible or absent momentum dependence within the energy dispersion, has garnered substantial attention within the domain of condensed matter physics[15–18]. Demonstrated to not only potentiate unconventional superconductivity[18–21] but also hold promise for engendering unconventional strong correlation effect[22]. The distinctive attribute of these bands resides in their copious density of states proximate to the Fermi energy, thereby augmenting electron pairing and elevating the critical temperature. The manifestation of superconductivity within flat bands has been observed across diverse systems, a prime example being twisted bilayer graphene[22]. On the other hand, the proximity of a Van Hove singularity (VHS) in the vicinity of the Fermi level, accompanied by a saddle point band structure, presents an alternate avenue to amplify instability and introducing novel quantum states such as superconductivity and charge density wave. For example, the divergence in the density of states, endowing the potential for heightened critical temperatures of superconductivity[23–27]; and the provocation of unconventional electron pairing via the anisotropic architecture and the coexistent presence of electrons and holes[28–30]. In this work we provide spectroscopic experimental substantiation for the coexistence of flat bands and Van Hove singularity within non-superconducting nitrogen doped lutetium hydride. These features hold the potential pathways towards the realization of high temperature superconductivity within hydrogen-rich materials in the future by tailoring their alignment with the Fermi energy[23].

Single crystals of nitrogen doped lutetium hydride have been successfully synthesized under ambient pressure conditions, exhibiting a distinctive dark-blue color as shown in Fig.1a. The crystal lattice host space group of Fm-3m with lattice parameter $a = b = c = 5.03$ Å proved by the outcomes of X-ray diffraction (XRD) experiments shown in Fig.1b. Comparative analysis validates the congruence of these crystals with those elucidated in the pertinent literatures[8,9]. Evidentiary support from Energy-Dispersive X-ray Spectroscopy (EDS) data establishes the incorporation of nitrogen within the material (Fig. 1c) same as that in Reference[9]. We conducted ultraviolet/X-ray photoelectron spectroscopy (UPS and XPS) experiments as depicted in Fig. 1d,e, which manifest peaks corresponding to lutetium (Lu) and hydrogen (H) elements. Notably, a subdivision of the Lu 4f core levels into a minimum of three distinct peaks was observed in both UPS and XPS spectra (Fig.1f). This splitting most probably serves as a testament to the diverse chemical states of Lu atoms engendered by the process of doping. Same as reported in references[9–12], the electrical transport data of our sample do not show any discernible superconducting phase transitions (Fig.1a). It is imperative to underscore that our data, similar as previous reports, encounter a limitation in elucidating the precise crystal structure. This inherent constraint stems from the intrinsic inability of XRD experiments to discern hydrogen atoms within the lattice. The XRD results of $LuH_2$ and $LuH_3$ show high similarity, which complicates the identification of their crystal structure. As posited in antecedent research, an inference was drawn towards the presence of $LuH_2$ or $LuH_3$ constituents, with a subset of hydrogen atoms potentially replaced by nitrogen counterparts[8,9,31].

The electronic structure of Lu-H-N is systematically studied by employing angle resolved photoemission spectroscopy (ARPES) technique. The resulting Fermi surface mapping, acquired with photon energy hv = 70 eV, is presented in Fig.2a. This depiction reveals the presence of multiple distinct Fermi surfaces. However, due to inherent limitations in sample single crystal quality, the discernment of periodic structural attributes within the Brillouin zones of momentum space proves to be a challenge. A series of spectra cuts, delineated by red dashed lines in Fig. 2a, are shown in Fig. 2b-h. All the cuts significantly show near-flat band features depicted by red-dot marks with light blue curves near the Fermi level. These features exhibit binding energies spanning the range of 0 to 200 meV. Notably, the distinctive character of flat bands is consistently evident across all the cuts. Particularly striking is the presence of well-defined dispersive bands behavior in each of the cuts. It serves as a compelling testament to the intrinsic single crystal quality of the sample under scrutiny. However, the sample comprises distinct single crystal domains, thereby introducing challenges in the alignment of the sample and the establishment of high symmetry points. Despite this inherent intricacy, it is essential to underscore that this variability does not undermine the central focus on the salient features of the flat bands. The energy distribution curves (EDCs) distinctly elucidate the presence of peaks corresponding to the flat band in immediate proximity to the Fermi level, as denoted by blue dashed lines in Fig. 2i. Concurrently, the salient dispersive character of the light bands is demarcated by green dashed lines, reaffirming their discernible presence within the electronic structure.

After a thorough deliberation of the distribution of the flat bands along in-plane $k$ positions that are parallel to the cleaved surface, a strategic tuning in photon energy was executed to investigate the out-of-plane $k$ information. The resulting spectra cuts, acquired utilizing photon energies 50, 60, 70, 80, 90, 100, 110 eV, are shown in Fig. 3a-d and Extended Data

Fig. S1. Evidently, the persistent presence of flat bands is discernible across all data sets on varying photon energies, denoted by red dashed lines. It is notable that the intensity of these flat bands exhibits a diminishing trend with increasing photon energy, notably becoming markedly attenuated for photon energies surpassing 100 eV as shown in Extended Data Fig. S1. This propensity can be attributed to the photoemission process and its sensitivity to the cross-sectional aspects. Remarkably, the selected range of photon energies cover more than a solitary Brillouin Zone (BZ), considering the lattice parameter 5.03 Å. This observation unveils a pivotal insight of the substantiation of the presence of flat bands within the entirety of the three-dimensional BZ. To mitigate potential surface effects inherent to ultraviolet measurements, a comprehensive exploration was undertaken via soft X-ray ARPES. Remarkably, the persistence of the flat bands is unequivocally evident across the entirety of the acquired data set. In this endeavor, photon energies spanning the range from 600 eV to 780 eV were selected, thereby encompassing a substantial portion of more than one Brillouin Zone along the $k_z$ direction. The regions of the flat bands are distinctly marked by red boxes, as depicted in Extended Data Fig. S2. Photon energy dependent density of states show the peaks of flat bands near Fermi level as shown in Extended Data Fig. S2k. This compelling observation serves as a decisive marker underscoring the inherent bulk nature of the flat bands.

After the comprehensive examination of the flat band phenomena, an additional noteworthy revelation emerges in the form of a VHS from dispersive bands proximate to the Fermi level, discernible specifically at a photon energy of $hv = 60$ eV. The detailed Fermi surface map, when subjected to magnified analysis, is showcased in Fig. 4a. Notably, a vertically oriented cut, as illustrated in Fig. 4b, delineates the presence of a hole-like band characterized by its uppermost band top situated around -0.1 eV. Conversely, a distinct trend is manifested within the horizontal cuts, as evident through three parallel cross-sectional slices depicted in Fig. 4c,d,e. These horizontal cuts consistently reveal the discernible manifestation of an electron-like band dispersion horizontally. This notable observation serves to underscore the existence of VHS electronic structures within the material, where both hole-like and electron-like bands concurrently contribute to the intricate electronic landscape along perpendicular directions. The 3D schematic band structure is shown in Fig. 4f revealing the VHS electronic structure near Fermi level.

In contrast to the behavior in pure $LuH_2$ and $LuH_3$ systems, where prior Density Functional Theory (DFT) calculations have indicated an absence of both flat bands and VHS proximate to the Fermi level[31,32], the situation is significantly altered by the introduction of nitrogen doping[33,34]. Specifically, our experimental findings decisively demonstrate the emergence of flat bands and VHS near the Fermi level, a trend that has been consistently postulated by previous computational investigations upon nitrogen incorporation into $LuH_3$ [33,34]. To further substantiate these empirical insights, we conducted ab initio calculations of the band structure for nitrogen doped $LuH_2$ and $LuH_3$. This computational endeavor involved the expansion of the unit cell and subsequent unfolding of the calculated bands into the original Brillouin Zone. The outcomes of these calculations for nitrogen doped $LuH_3$ with a composition of $Lu_8H_{23}N$ incontrovertibly reveal the presence of a conspicuous flat band proximal to the Fermi level, as well as potential VHS located at the X, or R point as shown in Extended Data Fig. S3. In intriguing contrast, our investigations into commensurate nitrogen doped $LuH_2$, with composition of $Lu_8H_{15}N$, do not evoke the emergence of any discernible flat bands in the vicinity of the Fermi level (Extended Data Fig. S4). In addition, the doping level is not uniform across the sample, leading to spatial variations in the local doping level. We investigate the effects of different doping levels on the electronic structure of $LuH_2$ with chemical formula $Lu_4H_7N$, $Lu_{32}H_{63}N$, and $Lu_{108}H_{215}N$ by calculating their unfolded band structures and displaying them in Extended Data Fig.

S5. We observe that higher doping level leads to more significant changes in the band structure. For $Lu_4H_7N$, the VHS at the W point is closer to the Fermi energy with a binding energy of about 0.3 eV, and a nearly flat band dispersion along the GX direction at around -0.5 eV. For $Lu_{32}H_{63}N$, we find two flat bands induced by nitrogen at -3 eV and -0.6 eV. For $Lu_{108}H_{215}N$, the doping level is too low to alter the band structure, which remains similar to that of the pristine parent compound $LuH_2$.

After investigation the calculation, we revisit the origin of the flat band observed in the experimental data. Our EDS analyses have yielded that the nitrogen ratio varies significantly across different spatial locations on the scale of tens of microns. This unexpectedly modest nitrogen doping level may introduces a non-negligible number of lattice defects, thereby possibly giving rise to flat band energy levels proximate to the Fermi level in the experiments. Intriguingly, another plausible scenario emerges wherein the nitrogen atoms establish intricate, larger-scale periodic arrangements within the lattice. Upon unfolding from this super lattice configuration, the potential arises for the generation of flat bands, akin to the fascinating phenomena encountered in the realm of magic-angle twisted bilayer graphene[22,35]. Future experiments are needed to elucidate the intricate details of the underlying crystal structure with heightened precision. Employing methodologies such as neutron deflection in the context of hydrogen isotope substitution samples offers a promising method toward the acquisition of a more comprehensive and accurate depiction of the crystal structure. We would like to emphasize that our current investigation does not offer conclusive evidence for the origin of the flat band.

In unraveling the intriguing emergence of VHS within the electronic structure, distinct pathways present themselves contingent upon the parent compound being either $LuH_3$ or $LuH_2$. (1) If the parent compound is $LuH_3$, our computational analyses underscore the feasibility of introducing VHS in proximity to the Fermi level through the imposition of a superlattice structure. This computational insight corroborates the possibility that the VHS observed in our experimental data may indeed be induced through such a mechanism[33]. (2) In the scenario where the parent compound is $LuH_2$, several discussions unfold. The observed band structure in our data deviates from that in the calculation, hinting at potential hole doping, possibly stemming from Lu vacancy defects, richer hydrogen, or nitrogen doping. Such hole doping might facilitate the tuning of the VHS at the W point in the calculation (Extended Data Fig. S5) to align more closely with the Fermi level. Alternatively, the VHS might be engendered by the band folding effect arising from a much higher superlattice structure formed by the nitrogen elements, akin to the phenomenon observed in the $LuH_3$ system[33]. Same as the flat band, our current investigation also does not offer conclusive evidence for the origin of the VHS.

The interplay of the flat band and VHS in the nitrogen-doped lutetium hydride system presents a nuanced perspective on their potential influence on superconductivity. Despite the proximity of their energy levels, it is crucial to underscore that both the flat band and VHS can only contribute significantly to superconductivity if they reside near the Fermi level and within the energy range encompassed by the superconducting gap. To harness their superconducting potential, a strategic shift of their energy levels towards the Fermi level is imperative. One plausible avenue for achieving this objective is through hole doping. Remarkably, a relatively modest doping requirement of around 6% additional hydrogen content might suffice to affect this energy tuning, thereby rendering the flat band and VHS conducive to participating in the superconducting mechanism. These propositions present fertile ground for future experimental and theoretical investigations, promising to unravel the intricate interplay between the flat band, VHS, and their contributions to possible superconductivity in the nitrogen-doped lutetium hydride system.

In summation, the experimental unveiling of the flat band and VHS in proximity to the Fermi level heralds a potentially transformative avenue towards the realization of high-temperature superconductivity within the nitrogen-doped lutetium hydride system. The intriguing prospects encompass not only elevated critical temperatures but also captivating manifestations of strong electron correlation, anisotropic symmetries, and enhanced electron-phonon coupling. As we await further confirmation through endeavors such as hole-doped single crystal synthesis and other innovative approaches, the potential for novel quantum states in nitrogen-doped lutetium hydride beckons a realm of fascinating possibilities and profound exploration.

**Materials and Methods**

For the preparation of LuHN sample, the metallic Lu flake with a thickness less than 1 mm and ammonia borane (AB) and $KN_3$ as the hydron and nitrogen sources were selected as raw materials. The stoichiometric Lu and AB, $KN_3$ were sealed in a quartz tube under argon in separated $Al_2O_3$ crucibles and then heated to 600 °C with a speed of 4°C per minute for 3h. After that, the sample was quenched to room temperature in 1 minute. The Lu-H-N single crystal with blue colour was obtained.

ARPES measurements were performed at the beamline UE112 PGM-2b-1[2] of BESSY (Berlin Electron Storage Ring Society for Synchrotron Radiation) synchrotron, at Bloch beam line of MAX-IV synchrotron, at ADRESS beamline of Paul Scherrer Institute. The energy and angular resolutions were set to ~30 meV and 0.1°, respectively. The samples for ARPES measurements were cleaved *in situ* and measured in a temperature range between 15 K and 30 K in a vacuum better than $5\times10^{-10}$ Torr.

The calculations were performed by using plane-wave density functional theory (DFT) with the Perdew-Burke-Ernzerhof[36] generalized gradient approximation (GGA) functional and the Quantum Espresso software package[37,38]. We adopted the pslibrary pseudopotentials for N, H and Lu[39] and set the wavefunction plane-wave cutoffs to 150 Ry and the charge density cutoff to 600 Ry. k-meshes are chosen corresponding to 0.1Å$^{-1}$ for Brillouin zone sampling. Hubbard U of 8.2 eV was applied for Lu as suggested in Ref.[33]. We studied both the pure $LuH_3$ and the nitrogen doped $LuH_{2.875}N_{0.125}$ with the $Fm\bar{3}m$ space group. We also investigated the pure $LuH_2$ with the $Fm\bar{3}m$ space group and the nitrogen doped $LuH_{1.875}N_{0.125}$ with the $F\bar{4}3m$ space group. VESTA code[40] was employed to visualize the crystal structures and the BandUP package[41,42] to unfold the bands for the doped systems.

**Acknowledgements**

We acknowledge Rui Lou, Sailong Ju, Fatima Alarab, Anna Sophia Hartl and Vladimir Strokov for help during the ARPES experiments. This work was supported by Research Grant Council (RGC) via the early career scheme (ECS) with grant number 21304023, via the Collaborative Research Fund (grant number C6033-22G), and the Collaborative Research Equipment Grant (grant number C1018-22E). J.Z.M. was partially supported by the National Natural Science Foundation of China (12104379), Guangdong Basic and Applied Basic Research Foundation (2021B1515130007). J.G.C. was supported by the National Natural Science Foundation of China (12025408, 11921004), the Beijing Natural Science Foundation (Z190008), the National Key R&D Program of China (2021YFA1400200), and the strategic Priority Research Program of CAS (XDB33000000).

**Author contributions**

J.Z.M. and Y.R. supervised this project. J.Z.M, X. L., Z.H.L. performed ARPES experiments with the help of S.Y.F., W.L.L., Z.Z. and B. M; J.Z., J.F.Z., G.D.W, L.C.S. and C.Q.J synthesized the single crystals of nitrogen-doped lutetium hydrides and conducted transport and XRD measurements. N.N.W., P.F.S. and J.G.C. synthesized the crystals of $LuH_2$ for preliminary study. Y.R. analyzed the crystal structure; J.Z.M, X.L., Z.H.L., and R.Z.L. plotted the figures with suggestions of M.N.; Z.H.L. and R.Z.L.

performed first-principles calculations of the band structure. All the authors contribute to the discussion of this projects. J.Z.M and X.L. wrote the manuscript.

**Competing interests**

The authors declare that they have no competing interests.

**Data and materials availability:** All data needed to evaluate the conclusions in the paper are present in the paper and/or the Supplementary Materials. Materials and additional data related to this paper may be requested from the authors.

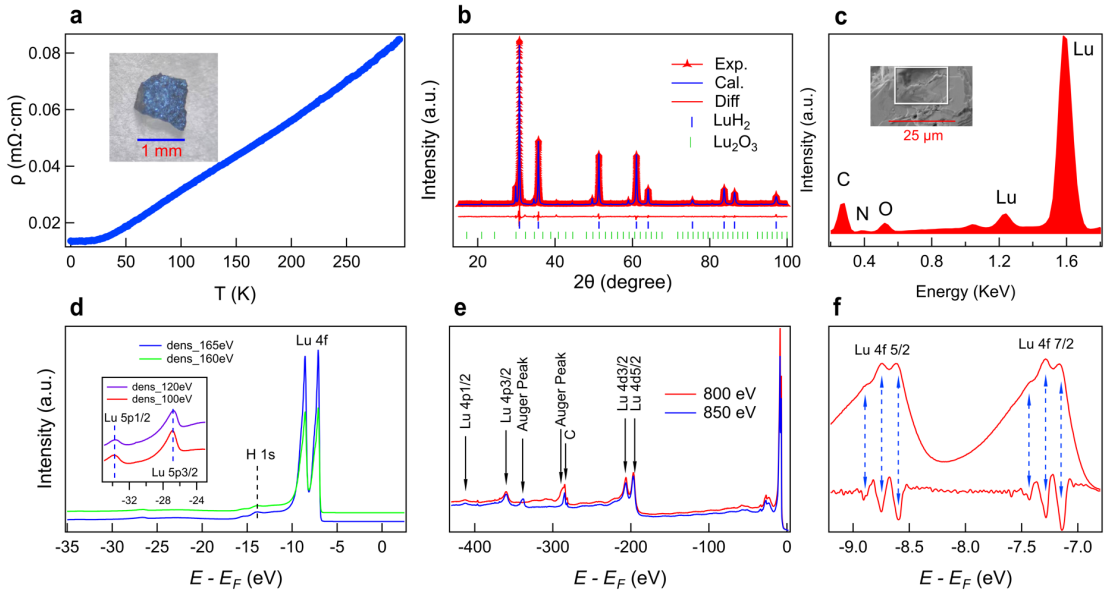

Fig.1. Crystal characterization of nitrogen doped lutetium hydride. a. 3D crystal structure, photos of Lu-H-N. b. The XRD curves plotted alongside a comparison with the default simulation database. c. EDS spectra showing the presence of nitrogen components within the sample. d,e. UPS and XPS spectra depicting core levels of Lu and H elements. f. Lu 4f orbital spectra exhibiting the distinctive splitting features.

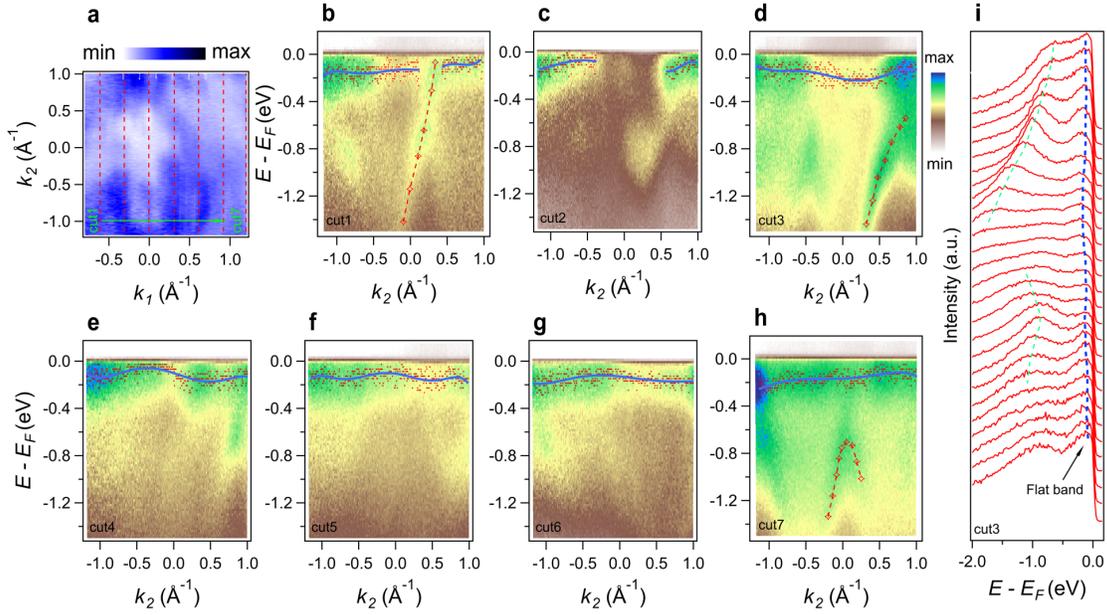

Fig.2. In plane flat band analysis. a. Fermi surface mapping acquired with photon energy $h\nu$ = 70 eV. b-h. A series of spectra cuts (designated as cut 1-7) showing the presence of flat band near the Fermi level, alongside dispersive bands. i. EDCs from cut3 revealing discernible peaks corresponding to the flat band near the Fermi level.

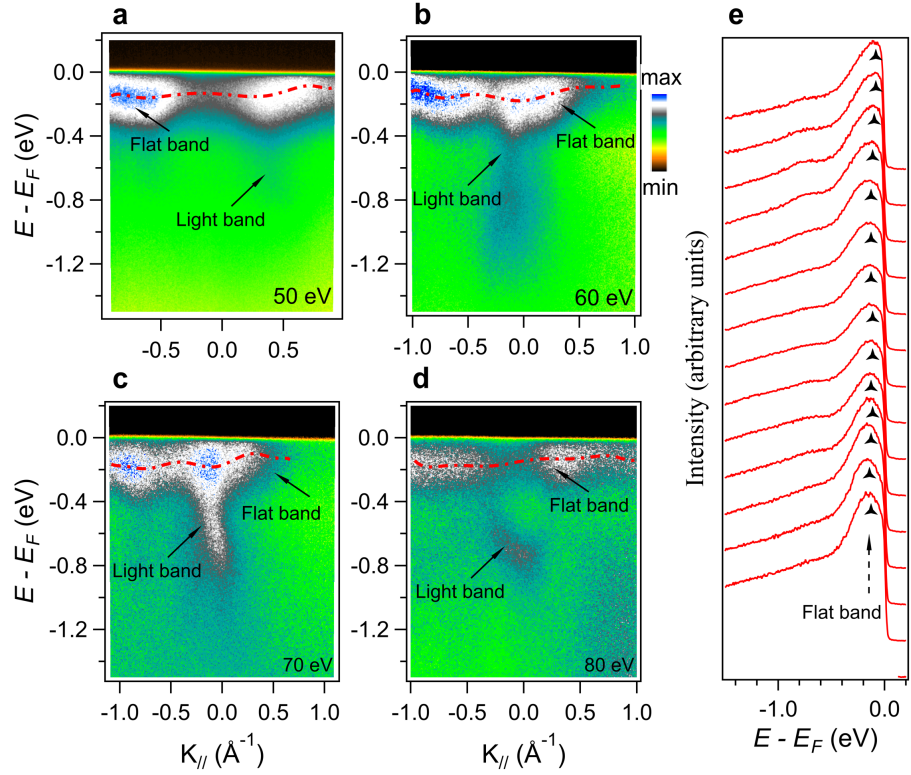

Fig.3. Out-of-plane flat band analysis. a-d. ARPES spectra acquired at varying photon energies, spanning from 50 eV to 80 eV respectively. e. EDCs extracted from the spectra of panel a, revealing distinct peaks corresponding to the flat band near the Fermi level.

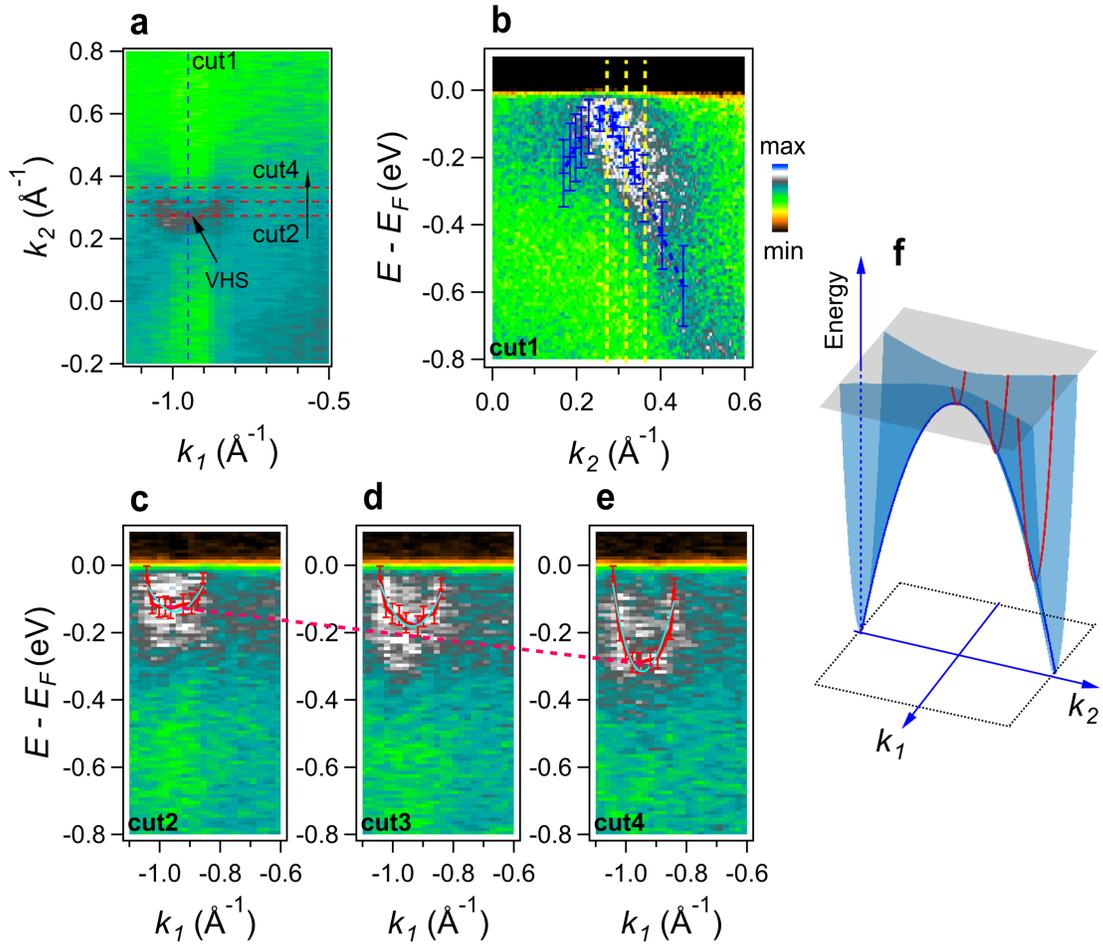

Fig. 4. Observation of VHS near the Fermi level. a. Fermi surface mapping acquired with photon energy hv = 60 eV. b-e, ARPES band spectra along cut 1-4 shown in panel a, elucidating the notable hole-like dispersion along k2 and electron-like dispersion along k1. f. Schematic of the VHS 3D band structure. The green band indicates the hole-like dispersion, and the red bands indicate the electron-like dispersion.

# Supplementary figures

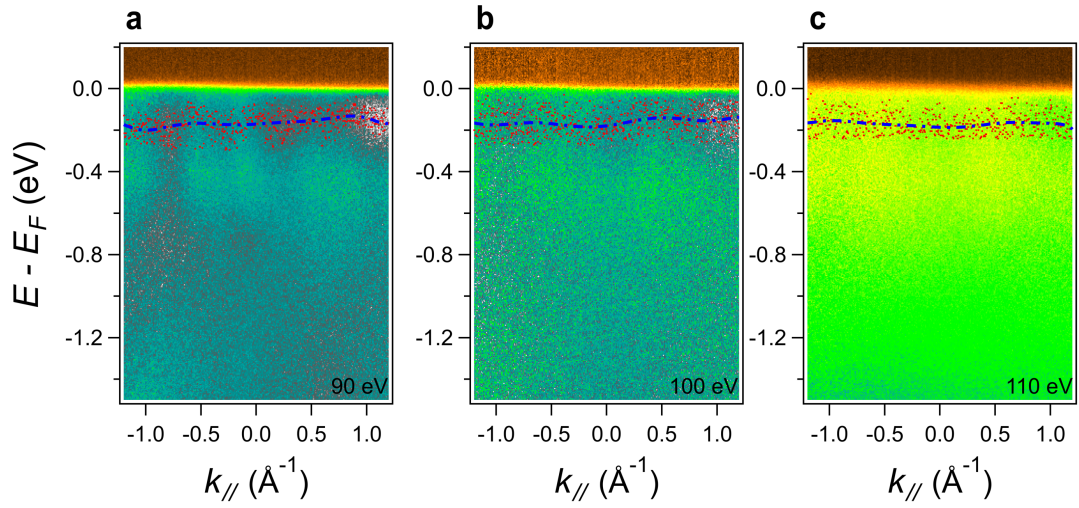

Fig. S1. Out-of-plane flat band analysis. a-c. ARPES spectra acquired at varying photon energies, spanning from 90 eV to 110 eV. The red dots indicate the flat band.

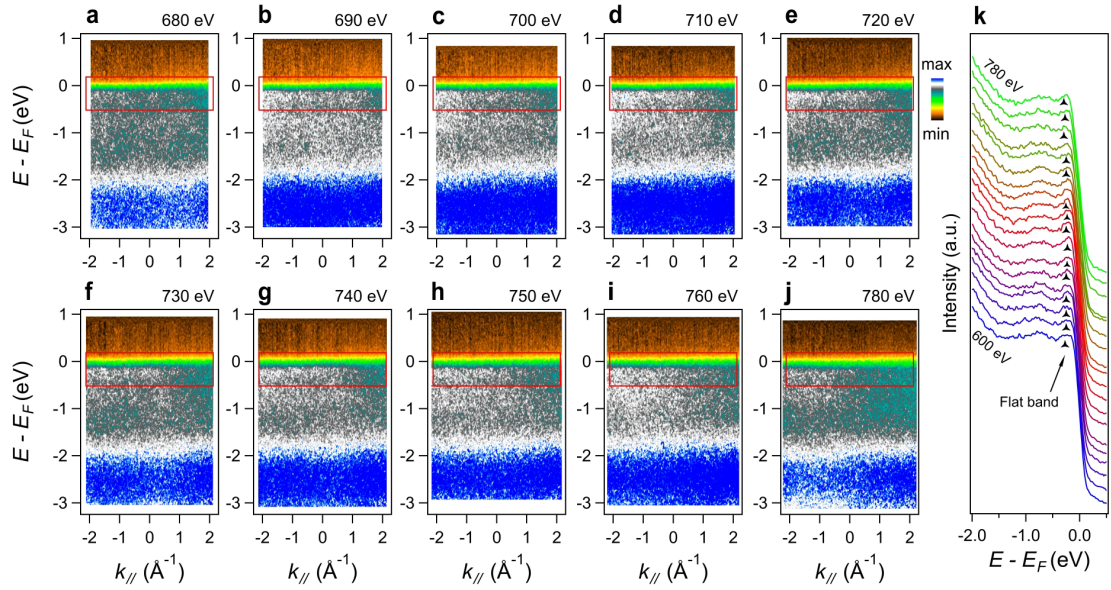

Fig. S2. Soft X-ray spectra and photon energy dependent density of states. a-j. ARPES spectra acquired over a range of photon energies, spanning from 680 eV to 780 eV. The red boxes show the region of interest with flat band near Fermi level. k. Photon energy dependent density of states, revealing the persistence of flat bands near the Fermi level.

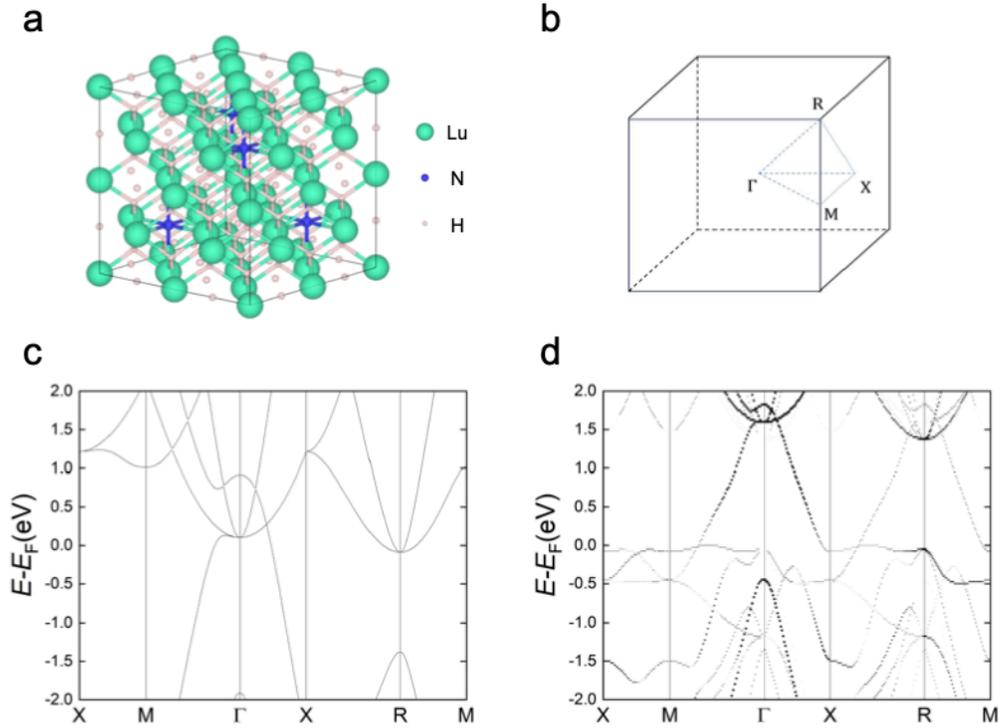

Fig. S3. Calculated band structure along high symmetry lines of parent and nitrogen doped LuH$_3$. To be comparable with ARPES experiments, we choose the path along in-plane directions in the BZ of original unit cell. a. Crystal structure of Lu$_8$H$_{23}$N. b. 3D unfolded BZ and the high symmetry points. c. Band structure of parent LuH$_3$. d. Unfolded band structure of Lu$_8$H$_{23}$N, revealing flat bands near the Fermi level along X-M-Γ, and X-R directions.

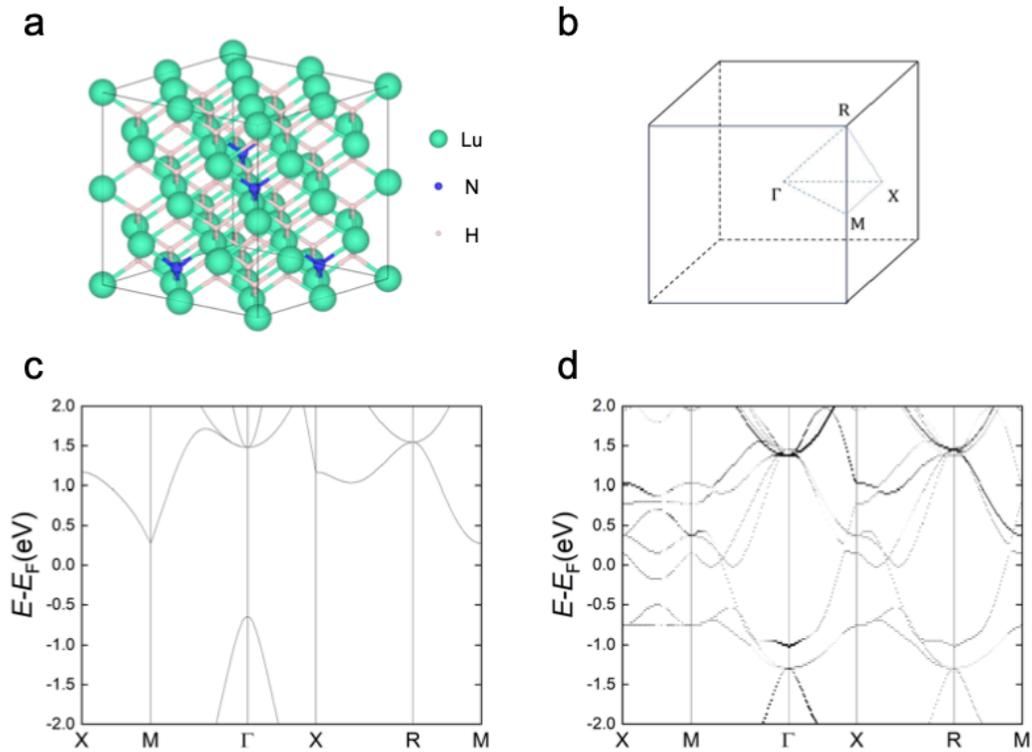

Fig. S4. Calculated band structure along high symmetry lines of parent and nitrogen doped LuH$_2$. To be comparable with ARPES experiments, we choose the path along in-plane directions in the BZ of original unit cell.   a. Crystal structure of Lu$_8$H$_{15}$N. b. 3D unfolded BZ and the high symmetry points. c. Band structure of parent LuH$_2$. Please be noted that along the chosen path there is gap between conduction band and valence bands. But it does not indicate an insulator state as conduction band crosses Fermi level along other path.   d. Unfolded band structure of Lu$_8$H$_{15}$N, revealing flat bands near the Fermi level.

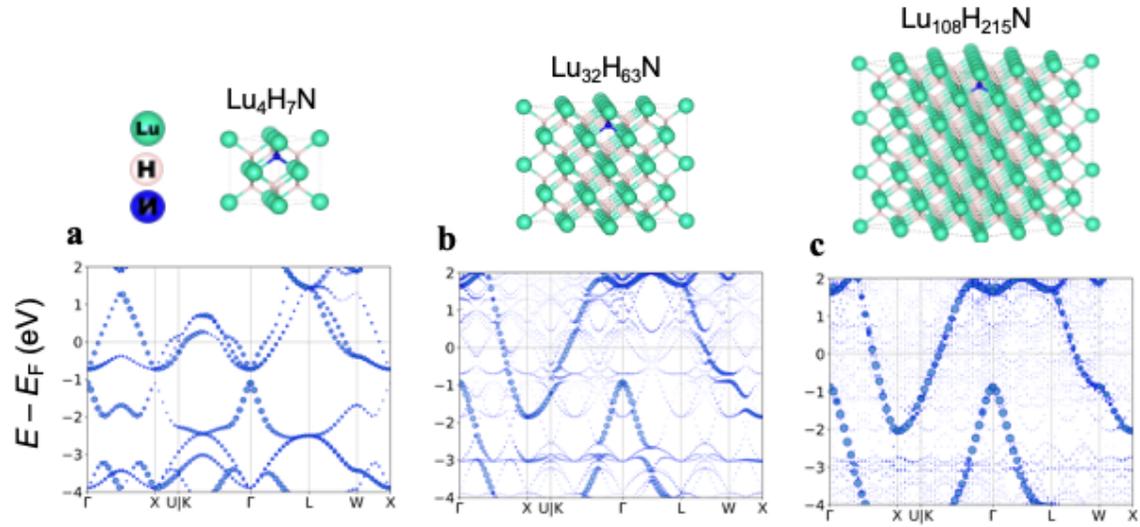

Fig. S5. Calculated band structure versus different nitrogen doping level of LuH$_2$. The bands are unfolded into the BZ of the original primitive cell. a. Unfolded band structure of Lu$_4$H$_7$N, showing near flat band along ΓX direction with energy spanning from -0.4 to -0.6 eV, and VHS at W point with energy around 0.4 eV below Fermi level. b. Unfolded band structure of Lu$_{32}$H$_{63}$N, revealing flat bands at around -3 eV and -0.6 eV. c. Unfolded band structure of Lu$_{108}$H$_{215}$N, indicating almost same band dispersion as the original LuH$_2$.